\documentclass[12pt,preprint]{aastex}

\shorttitle{}
\shortauthors{Nesvorn\'y et al.}

\begin{document}
\baselineskip 19.pt

\title{Dynamical Implantation of Blue Binaries in the \\Cold Classical Kuiper Belt}

\author{David Nesvorn\'y$^1$, David Vokrouhlick\'y$^{2}$, Wesley C. Fraser$^3$}
\affil{(1) Department of Space Studies, Southwest Research Institute,\\
1050 Walnut St., Suite 300, Boulder, CO, 80302, USA}
\affil{(2) Institute of Astronomy, Charles University,\\ 
V Hole\v{s}ovi\v{c}k\'ach 2, CZ--18000 Prague 8, Czech Republic}
\affil{(3) Planetary Science Institute,\\ 1700 East Fort Lowell, Suite 106, Tucson, AZ 85719,
USA}
\begin{abstract} 
Colors and binarity provide important constraints on the Kuiper belt formation. The cold classical 
objects at radial distance $r=42$--47 au from the Sun are predominantly very red (spectral 
slope $s>17$\%) and often exist as equal-size binaries ($\sim 30$\% observed binary fraction). 
This has been taken as evidence for the in-situ formation of cold classicals. 
Interestingly, a small fraction ($\sim 10$\%) of cold classicals is less red with 
$s<17$\%, and these ``blue'' bodies are often found in wide binaries. Here we study the dynamical 
implantation of blue binaries from $r<42$ au. We find that they can be implanted into the 
cold classical belt from a wide range of initial radial distances, but the survival of the 
widest blue binaries -- 2001 QW322 and 2003 UN284 -- implies formation at $r>30$ au.
This would be consistent with the hypothesized less-red to very-red transition at $30<r<40$ au.
For any reasonable choice of parameters (Neptune's migration history, initial disk profile, 
etc.), however, our model predicts a predominance of blue singles, rather than blue binaries, 
which contradicts existing observations. We suggest that wide blue binaries formed in situ
at $r=42$--47 au and their color reflects early formation in a protoplanetary gas disk. The 
predominantly VR colors of cold classicals may be related to the production of methanol and 
other hydrocarbons during the late stages of the disk, when the temperature at 45 au dropped 
to $\simeq 20$ K and carbon monoxide was hydrogenated.
\end{abstract}


\section{Introduction}

The Kuiper belt holds important clues about the Solar System formation. Neptune's early 
migration (Fern\'andez \& Ip 1981, Malhotra 1993), in particular, is thought to be responsible 
for the complex orbital distribution of Kuiper belt objects (KBOs). Observations and dynamical 
modeling of KBOs (see Morbidelli \& Nesvorn\'y (2020) and Gladman \& Volk (2021) for recent 
reviews) help us to understand: (i) the mass and radial structure of the original outer disk 
of planetesimals, and (ii) the nature and timescale of Neptune's migration (e.g., Hahn \& 
Malhotra 2005, Levison et al. 2008, Nesvorn\'y \& Vokrouhlick\'y 2016). For example, it is 
inferred that Neptune relatively slowly migrated (one $e$-fold $\gtrsim 10$ Myr, Nesvorn\'y 
2015a) on a modestly eccentric orbit ($e \sim 0.03$--0.1; Dawson \& Murray-Clay 2012, 
Nesvorn\'y 2021). Neptune's orbit could have been excited by a dynamical instability 
(Tsiganis et al. 2005, Nesvorn\'y \& Morbidelli 2012). 

KBO colors and binarity provide crucial constraints. The color distribution of small KBOs 
is bimodal with less red (LR) and very red (VR) objects observed, in different proportions, 
in all dynamical categories (e.g., Wong \& Brown 2017). Complex organics and surface weathering 
are thought to be responsible for different colors. The VR bodies are believed to have formed 
beyond an iceline of some volatile such as methanol or ammonia (Brown et al. 2011). The 
exact location of the LR-to-VR transition, $r^*$, is unknown. The dynamically cold KBOs at 
$r=42$--45 au are predominantly VR which indicates $r^*<42$ au. If $r^*<30$ au, however, 
the dynamically hot populations such as Plutinos, hot classicals (HCs) and scattered disk 
objects (SDOs) would presumably have a much larger proportion of VR objects (Nesvorn\'y et al. 
2020, Ali-Dib et al. 2021). This indicates $30<r^*<42$ au.  

Observations reveal an unusually high fraction of binaries among cold classicals (CCs; 
$\sim 30$\%, Noll et al. 2020). The CC binaries have nearly equal-size components ($R_2/R_1>0.5$, 
where $R_1$ and $R_2$ are the primary and secondary radii) and wide separations ($a_{\rm b}/(R_1+R_2) 
\sim 10$--1000, where $a_{\rm b}$ is the binary semimajor axis). They presumably formed during 
the formation of KBOs themselves, by the streaming instability (SI; Youdin \& Goodman 2005) 
and subsequent gravitational collapse (Nesvorn\'y et al. 2010, 2019). Their survival in different 
KBO categories reflects how much these populations were affected by collisions and dynamical 
processes (Petit \& Mousis 2004, Parker \& Kavelaars 2010). For example, only a very small 
fraction of wide binaries is expected to survive the dynamical implantation from $r<30$ au 
(most become unbound during encounters to Neptune; Parker \& Kavelaars 2010, Nesvorn\'y \& 
Vokrouhlick\'y 2019).

Fraser et al. (2017, 2021) reported detection of several ``blue'' binaries in the CC population 
(i.e., LR colored; spectral slope $s<17$\% as defined in Fraser et al. 2017; Table~1).  
This is surprising because if these binaries formed -- like most CCs -- beyond 42 au, they should 
be VR -- like most CCs. They could not have formed in the massive disk below 30 au, which is 
the source of most dynamically {\it hot} KBOs, presumably because they would not, especially 
the wide ones, survive (Parker \& Kavelaars 2010). It has been suggested (Fraser et al. 2017, 
2021) that the blue binaries formed at $r \sim 38$ au, from where that were pushed -- by migrating 
Neptune's 2:1 resonance -- to $r>42$ au. As this implantation mechanism does not involve 
gravitational scattering by Neptune, binaries are retained. This interpretation would indicate 
the LR-to-VR transition at $r^* \simeq 38$--42 au. 

Here we re-examine the implantation of blue binaries from $r<42$ au to $r>42$ au. The best dynamical 
models from Nesvorn\'y et al. (2020) are used to determine the implantation probability to an
orbit with the semimajor axis $42<a<47$ au, inclination $i<5^\circ$, and stable perihelion distance 
$q>36$ au. The dynamical survival of binaries during planetary encounters is evaluated as a 
function of the mutual separation of binary components. We find that the widest blue binaries -- 
2001 QW322 and 2003 UN284 -- would not survive if they formed below 30 au ($\lesssim1$\% survival 
probability). Their survival becomes progressively more likely with the increasing 
starting radius: $\sim$10--25\% for $30<r<36$ au, $\sim$40--50\% for $36<r<38$ au and 
$\sim$100\% for $r>38$ au. We discuss the implication of these, and other results obtained here, 
for blue binaries, the LR-to-VR transition in the original disk, and the interpretation 
of KBO colors in general.

\section{Dynamical Effect of Planetary Encounters}

We make use of two Kuiper belt simulations published in Nesvorn\'y et al. (2020); they are referred to 
as 10/30 and 30/100 here (Table 2). See that work for 
the description of the integration method, planet migration, initial orbital distribution of 
disk planetesimals, and comparison of the results with the orbital structure of the Kuiper 
belt. A shared property of the selected runs is that Neptune migrates outward by scattering 
planetesimals. Planetesimals are initially distributed in a disk extending from 
just beyond the initial orbit of Neptune at 22 au to $>50$ au. The radial distribution of 
planetesimals is assumed to have a drop at $\simeq30$ au, with the massive inner disk and 
low-mass outer extension, or a smooth profile with the surface density exponentially decreasing 
with distance. Either of these initial distributions can match existing observational 
constraints (Nesvorn\'y et al. 2020). The simulations were performed with the symplectic $N$-body 
integrator known as {\it Swift} (Levison \& Duncan 1994). 

All encounters of planetesimals with planets were recorded. This was 
done by monitoring the distance of each planetesimal from Jupiter, Saturn, Uranus and Neptune, 
and recording every instance when the distance dropped below 0.5 $R_{{\rm Hill},j}$, where 
$R_{{\rm Hill},j}$ are the Hill radii of planets ($j=5$ to 8 from Jupiter to Neptune, 
the inner planets were not included). We verified 
that the results do not change when more distant encounters are accounted for. The disk 
planetesimals that ended up on orbits with with $42<a<47$ au, $q>36$ au and $i<5^\circ$ at the 
end of the simulations ($t=4.5$ Gyr) were selected and used for the analysis presented here.
See Nesvorn\'y \& Vokrouhlick\'y (2019) for the binary survival during implantation into the 
dynamically hot populations.

Each selected planetesimal was assumed to be a binary object. We considered a range of binary 
separations, $1<a_{\rm b}/R_{\rm b}<2000$, where $R_{\rm b}=(R_1^3+R_2^3)^{1/3}$ (Agnor 
\& Hamilton 2006), initially circular orbits (binary orbit eccentricity $e_{\rm b}=0$), and 
a random distribution of binary inclinations ($i_{\rm b}$). Ten clones with different binary 
inclinations were assigned to each selected planetesimal to increase the statistic. 
Each binary was evolved through all recorded planetary encounters. We used the Bulirsch-Stoer (B-S) 
$N$-body integrator that we adapted from Numerical Recipes (Press et al. 1992). The center of mass 
of the binary was first integrated backward from the time of the closest approach to 
3~$R_{{\rm Hill},j}$. It was then replaced by the actual binary and integrated forward through the 
encounter until the planetocentric distance of the binary exceeded 3~$R_{{\rm Hill},j}$. The final binary 
orbit was used as an initial orbit for the next encounter and the algorithm was repeated over 
all encounters. 

The B-S code monitored collisions between binary components. If a collision occurred, the integration 
was stopped and the impact speed and angle were recorded. A fraction of binaries became unbound. 
For the surviving binaries, we recorded the final values of $a_{\rm b}$, $e_{\rm b}$ and $i_{\rm b}$, 
which were then used to evaluate the overall change of orbits. The statistics of surviving binaries 
was used to compute the binary survival probability, $p_{\rm b}$, as a function of $a_{\rm b}/R_{\rm b}$.
  
\section{Implantation Probability}

\begin{figure}
\epsscale{1.0}
\plotone{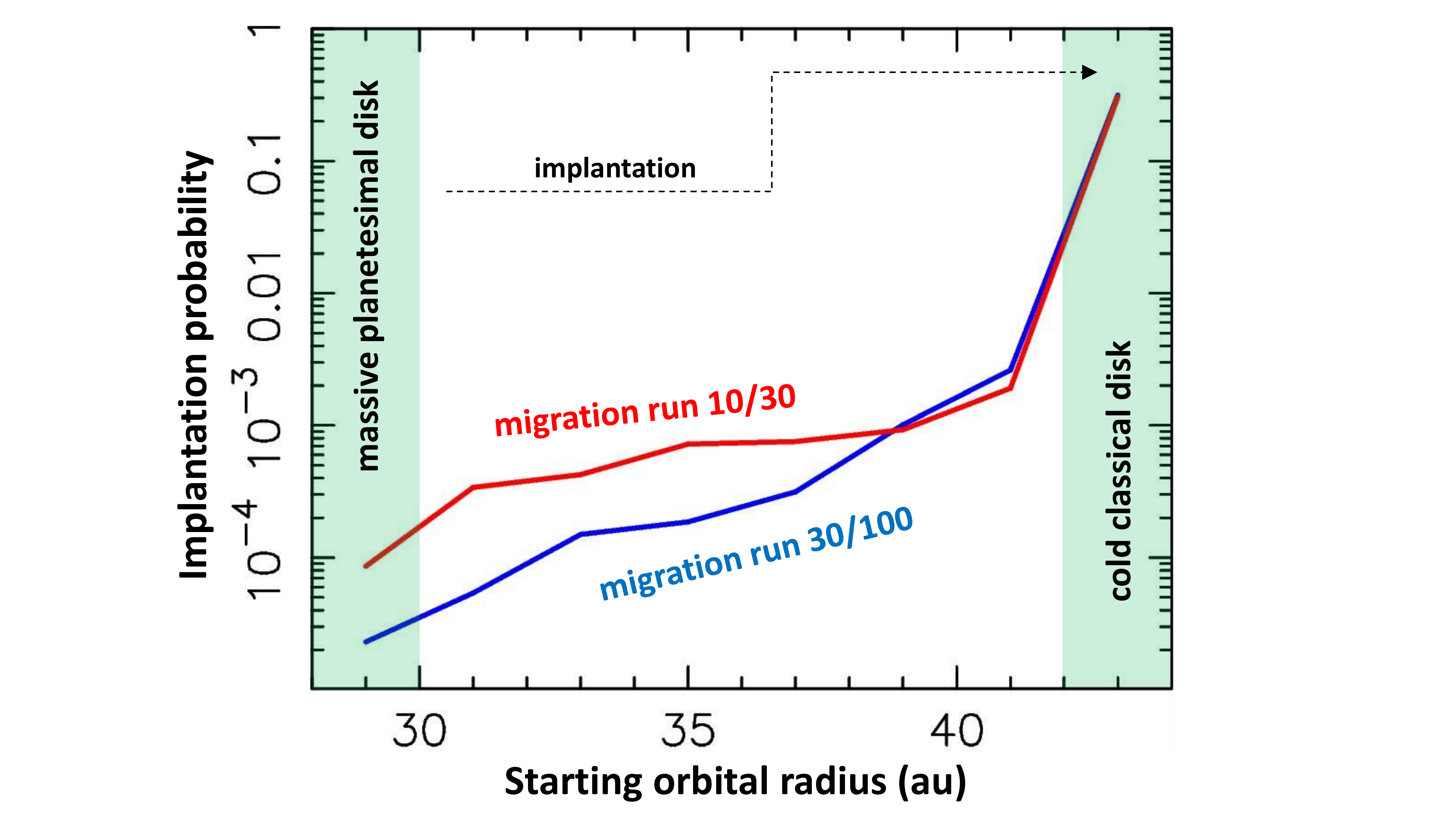}
\caption{The implantation probability to an orbit with $42<a<47$ au, $i<5^\circ$ and $q>36$ au as a function
of the starting orbital radius. The implantation probability for $r<30$ au is $\lesssim 10^{-4}$. For $r<38$ au, 
the implantation probabilities are higher in the 10/30 simulation (faster migration of Neptune)
than in 30/100 (slower migration). Bodies starting in a cold disk at $42<r<47$ au have a 30--40\% chance 
to end up as cold classicals (Nesvorn\'y 2015b).}
\label{pops}
\end{figure}

Figure \ref{pops} shows the implantation probability onto an orbit with $42<a<47$ au, $i<5^\circ$ and 
$q>36$ au from different parts of the original planetesimal disk. The implantation probability
is computed as $p_j = n_j/N_j$, where $N_j$ is the number of planetesimals starting with 
$r_{j-1}<r<r_j$ in our simulations, and $n_j$ is the number of planetesimals starting with 
$r_{j-1}<r<r_j$ {\it and} ending with $42<a<47$ au, $i<5^\circ$ and $q>36$ au. Probability $p_j$
does not account for the availability of planetesimals in different zones. We used eight
zones in total with $r_0=24$ au $r_j=30+2(j-1)$ au for $j=1,\dots,7$, and $r_8=47$ au. Note that 
the first zone ($j=1$) corresponds to the full radial span of the inner massive disk, here assumed 
to be 6 au wide, and the eighth zone ($j=8$) is the orbital radius of the cold classical disk; all 
other zones are 2 au wide.

The implantation probability increases with the initial radial distance, from $p_2 \sim 10^{-4}$ 
for $r \simeq 31$ au to $p_7 \simeq(2$--$3)\times 10^{-3}$ for $r \simeq 41$ au. It is apparently easier 
to implant bodies in the cold classical belt if they start radially closer to the belt. 
For the eighth zone, where bodies already start in the CC region, $p_8$ is the probability 
of survival on a CC orbit. We find, consistently with the previous results (Nesvorn\'y 
2015b), that 30-40\% of planetesimals starting in the cold classical belt survive ($p_8 
\simeq 0.3$--0.4). The implantation probabilities from the massive disk below 
30 au are low: $p_1 \simeq 8 \times 10^{-5}$ for 10/30 and $p_1 \simeq 2 \times 10^{-5}$ for 
30/100. The probability profile less steeply raises with the initial orbital radius if Neptune's 
migration is faster, and is practically flat between 30 and 42 au for the 10/30 simulation.  
 
We now multiply the implantation probability by the number of planetesimals initially available 
in each zone, $\eta_j$. We consider two cases from Nesvorn\'y et al. (2020): the (1) truncated 
power-law profile (the surface density $\Sigma \propto 1/r$, truncated at 30 au), and (2) 
exponential profile ($\Sigma \propto \exp [(r-r_0)/\Delta r]/r$, where $r_0$ denotes the inner 
edge radius at 24 au, and $\Delta r$ au is one $e$-fold, no outer truncation here). 
For the truncated power-law profile, the change of the surface density at $r=30$ au is parametrized 
by the contrast parameter, $c$, which is simply the ratio of surface densities on either side of 30 au. 
Here we use $\Delta r = 2.5$ au and $c=1000$. Both these cases were shown to correctly reproduce 
the orbital and color distribution of KBOs (Nesvorn\'y et al. 2020). 

We define $n_{{\rm CC},j} = p_j \eta_j$, where $n_{{\rm CC},j}$ represents the actual number of planetesimals 
implanted onto the CC orbits from different zones, and normalize $n_{{\rm CC},j}$ such that 
$n_{{\rm CC},8}=10^4$. This means that we expect $10^4$ planetesimals to survive in the CC region 
($\sim 1/3$ of the original population; Fig. \ref{pops}). For comparison, Kavelaars et al. (2021) 
estimated from the Outer Solar System Origins Survey (OSSOS; Bannister et al. 2018) observations that there 
are $\simeq$5,000--15,000 CCs with diameter $D>100$ km (the range given here reflects the uncertain 
albedo of CCs, $p_{\rm V}=0.05$--0.2).

\begin{figure}
\epsscale{1.0}
\plotone{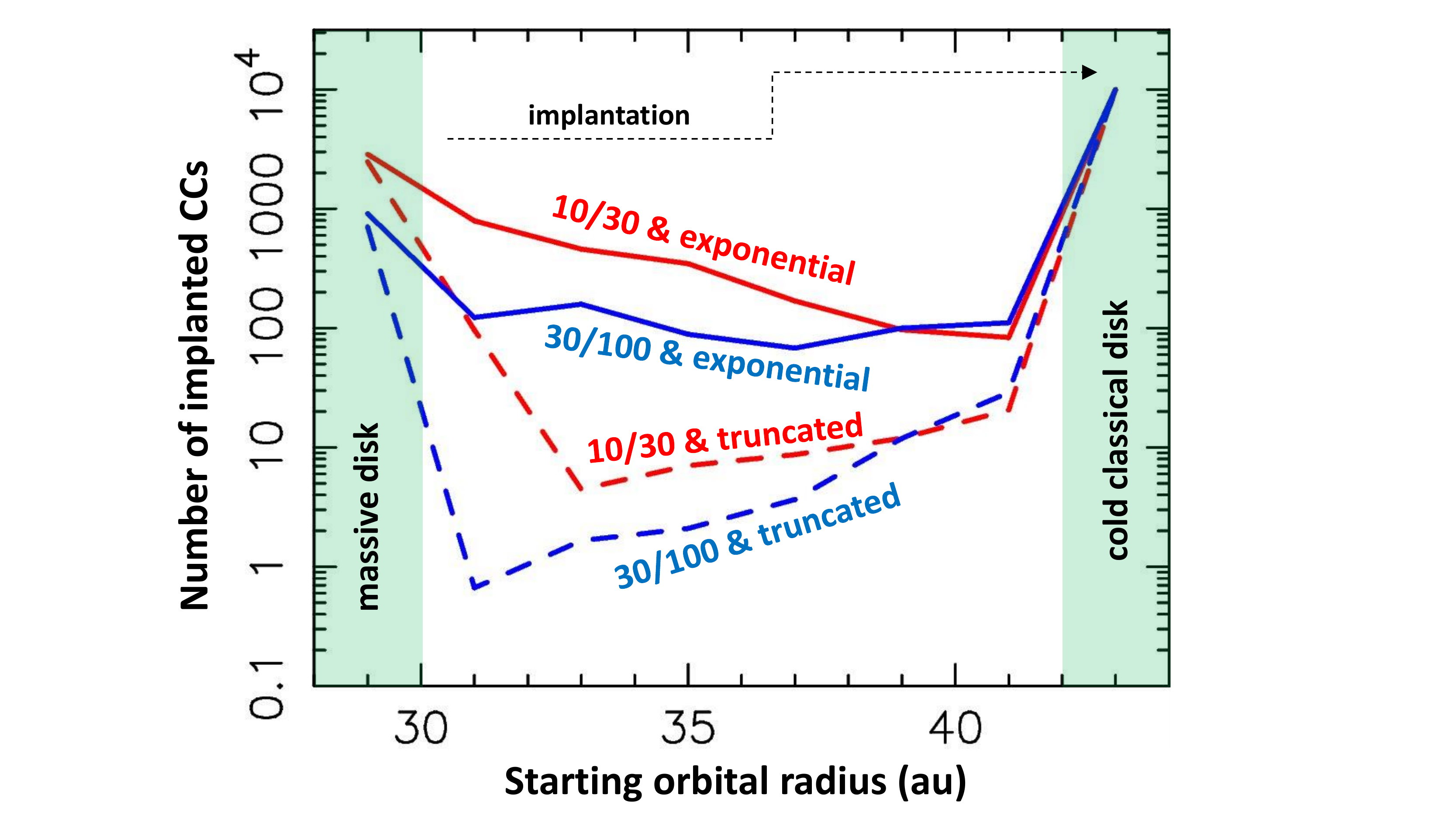}
\caption{The number of bodies ($n_{{\rm CC},j}$) implanted from different starting radii on orbits with 
$42<a<47$ au, $i<5^\circ$ and $q>36$ au. We normalize $n_{{\rm CC},j}$ such that $n_{{\rm CC},8}=10^4$. The 
blue and red lines show the results for faster (10/30) and slower (30/100) migration of Neptune, respectively. The solid and dashed 
lines are the exponential and truncated disks, respectively.}
\label{weis}
\end{figure}

With the truncated power-law profile, the 30--42 au region does not contribute much to the population 
of CCs, for either migration run considered here (Fig. \ref{weis}). In this case, we would expect 
$\sim 100$ bodies to be implanted from $r=30$--42 au, only $\sim 1$\% of today's CC population.
For the inner part of the planetesimal disk, $r<30$ au, we find $n_{{\rm CC},1} \sim 2500$ for 10/30 and 
$n_{{\rm CC},1} \sim 800$ for 30/100. This would represent $\sim 8$--25\% of the CC population and imply 
that nearly all blue CC binaries would need to be implanted from $r<30$ au. This is difficult to reconcile 
with observations (Fraser et al. 2017, 2021), because the wide binaries such as 2001 QW322 and 2003 
UN284 would not dynamically survive (Sect. 4; Nesvorn\'y \& Vokrouhlick\'y 2019). The population
of blue cold classicals would be dominated by singles in this case -- the opposite to what is observed. 

The results with the exponential disk profile are somewhat more plausible. Here, the contribution
of the $r<30$ au and $30<r<42$ au source regions would be similar: 15--30\% in 10/30 and $\sim6$--10\% 
in 30/100 (percentages given relative to the present CC population; Kavelaars et al. 2021). This 
would also produce roughly the right proportion of LR/VR CCs assuming that $30 < r^* \lesssim 40$ au 
for 10/30 or $35 < r^* \lesssim 40$ 
au for 30/100 (Nesvorn\'y et al. 2020). In the next section, we discuss the dynamical survival of 
binaries starting in different zones.  
 
\section{Binary Survival}

\begin{figure}
\epsscale{1.1}
\hspace*{-1.cm}\plotone{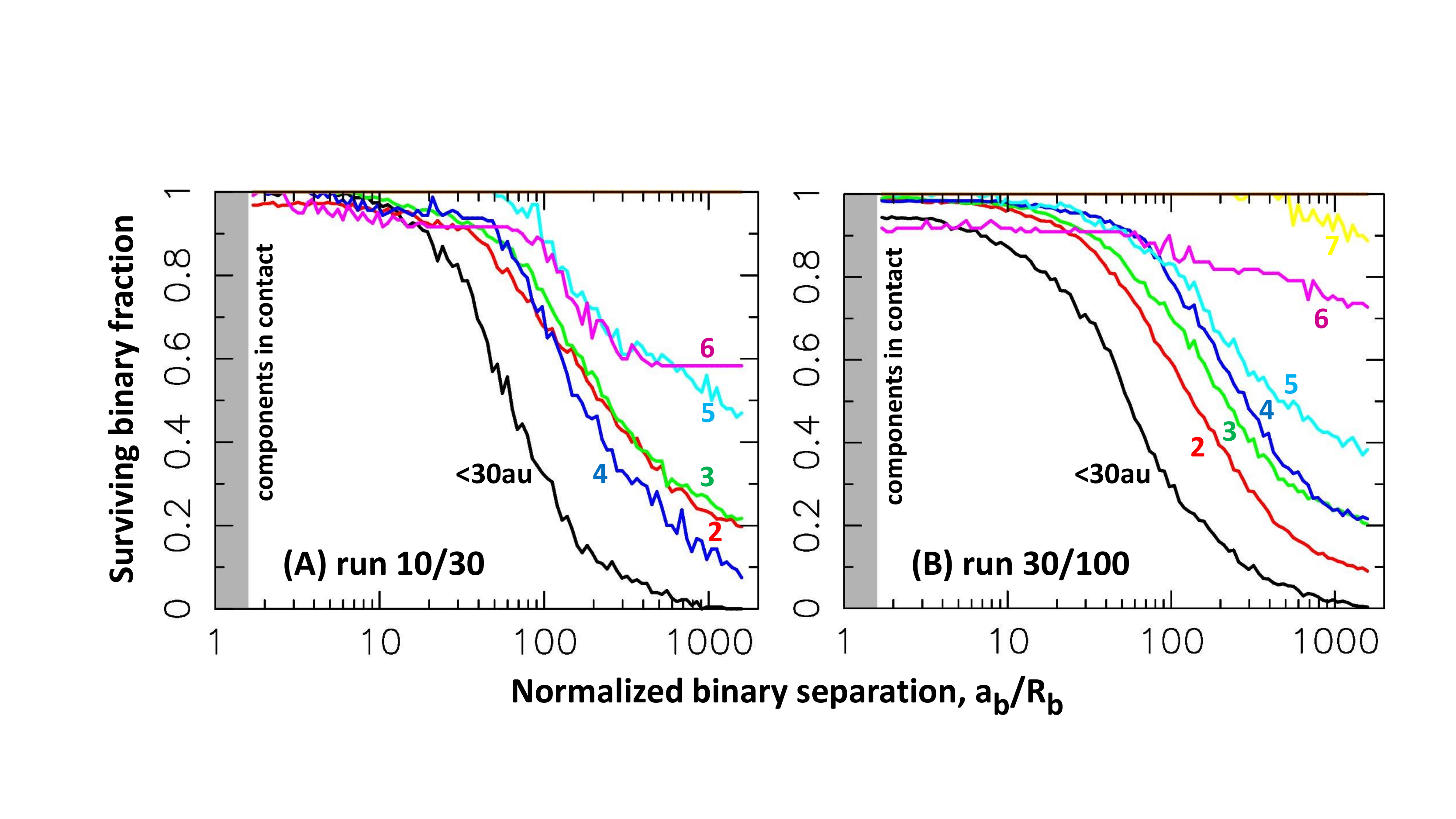}\vspace*{-1.cm}
\caption{The dynamical survival of binaries as function of their starting orbital radius and separation.
The surviving fraction of wide binaries is relatively low because they are more susceptible to gravitational
perturbations during planetary encounters. The surviving fraction increases with the starting orbital radius
because fewer encounters happen during the implantation stage for these initially more distant bodies. The
yellow line corresponding to 40--42 au (zone 7) is missing in the left panel, because we did not have
sufficient statistics in this case.}
\label{bins}
\end{figure}

Figure \ref{bins} shows how the survival probability changes with the starting heliocentric distance 
and binary separation. We find that the binary survival depends on $a_{\rm b}/R_{\rm b}$, where 
$R_{\rm b}=(R_1^3+R_2^3)^{1/3}$, 
and not on $a_{\rm b}$, $R_1$ and $R_2$ individually. This is a consequence of the binary dissociation 
condition described in Agnor \& Hamilton (2006). A binary with the total mass $m_{\rm b}=m_1+m_2$ 
can become unbound when the planetocentric Hill radius of the binary, $r_{\rm Hill, b}=q(m_{\rm b}/3 
m_{\rm pl})^{1/3}$, where $q$ is the distance of the closest approach and $m_{\rm pl}$ is the planet 
mass, becomes smaller that the binary separation; that is $r_{\rm Hill, b} < a_{\rm b}$. This 
condition yields
\begin{equation}
{a_{\rm b} \over R_{\rm b}} > {1 \over 3^{1/3}} \left({\rho \over \rho_{\rm pl}}\right)^{\!\!\!1/3} 
\!\! \left( {q \over R_{\rm pl}} \right ) \ ,
\end{equation} 
where $R_{\rm pl}$ and $\rho_{\rm pl}$ are the planet radius and density. Here we assumed that 
the primary and secondary components of binaries have the same density, $\rho$. For exactly 
equal-size binaries with $R_1=R_2$, $R_{\rm b}=(R_1+R_2)/2^{2/3}$.  

The encounter statistics changes with the starting heliocentric distance and this projects into the 
expectation for binary survival. The survival probability is also a strong function of the 
binary separation (Fig. \ref{bins}).\footnote{The results shown in Fig. \ref{bins} were 
computed assuming the bulk density $\rho=1$ g cm$^{-3}$. According to Eq. (1), the critical 
semimajor axis scales with $\rho^{1/3}$. Therefore, the surviving fraction curve shown in Fig. 
\ref{bins} would shift left by a multiplication factor of 0.79 for $\rho=0.5$ g cm$^{-3}$ and 
right by a multiplication factor of 1.26 for $\rho=2.0$ g cm$^{-3}$. We confirmed this by 
simulating cases with different densities.} For the ease of presentation, we consider three 
types of binaries: tight with $a_{\rm b}/R_{\rm b} \sim 10$, wide with $a_{\rm b}/R_{\rm b} \sim 
100$, and extreme with $a_{\rm b}/R_{\rm b} \sim 1000$. For reference, one tight, five wide and 
two extreme blue CC binaries are currently known (Table 1). 

\begin{figure}
\epsscale{1.1}
\hspace*{-1.cm}\plotone{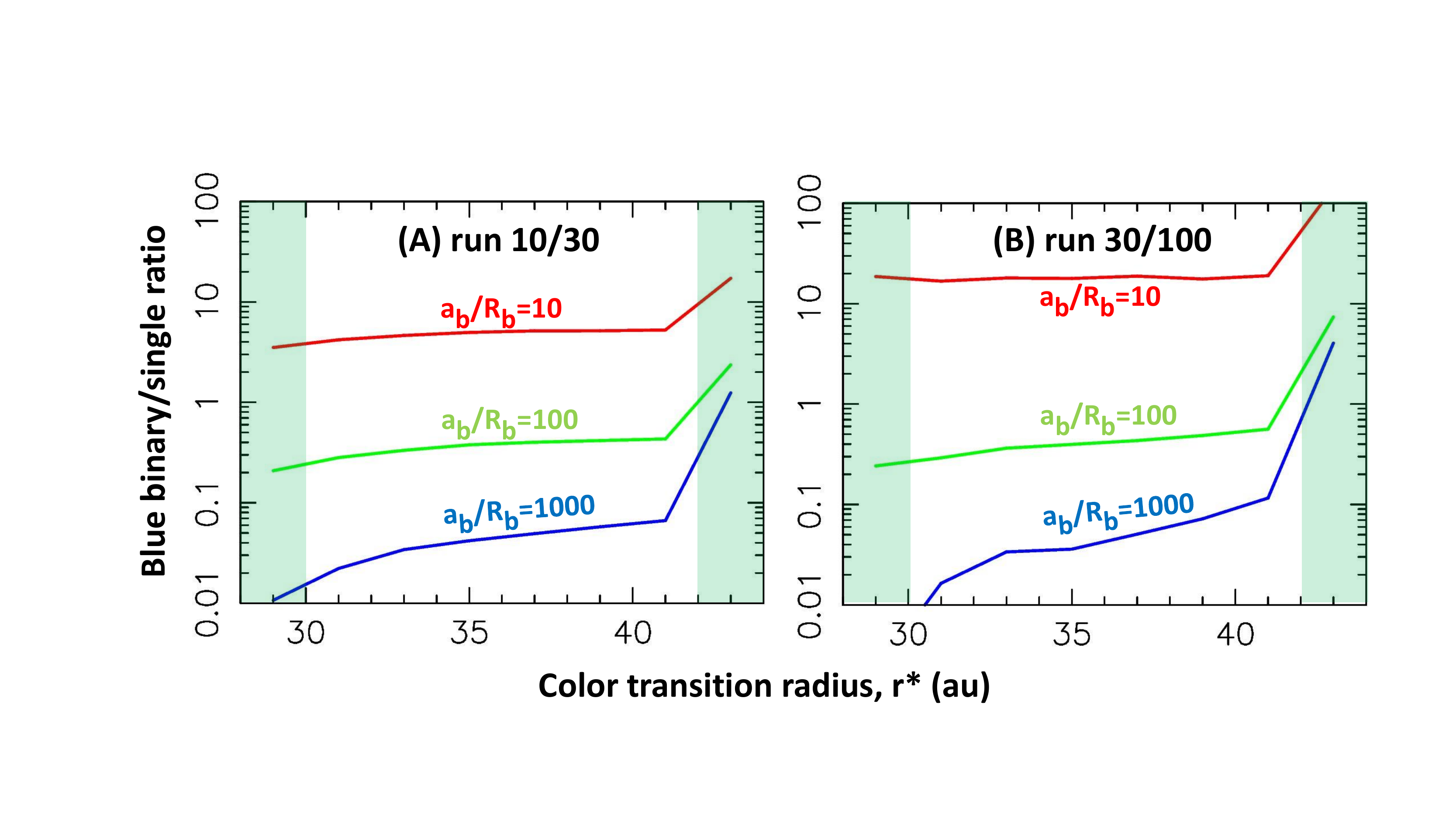}\vspace*{-1.cm}
\caption{The ratio of blue binaries to blue singles ($R_{\rm b/s}$) expected from the implantation process 
as a function of the LR-to-VR transition radius, $r^*$. In each case, we assumed a monodisperse distribution of binary 
separations: $a_{\rm b}/R_{\rm b}=10$ (red), 100 (green), and 1000 (blue). The initial binary fraction throughout 
the original planetesimal disk was assumed to be 100\% (no initial singles). We evaluate the proportion 
of surviving and dissolved binaries in each case and plot the expected blue binary-to-single ratio in the 
CC population. The results for the exponential disk profile are shown here. The profiles for the truncated disk,
where the implanted population is dominated by bodies starting with $r<30$ au, are less plausible.}
\label{ratio}
\end{figure}

The tight binaries with small separations are strongly bound together and have relatively high 
survival probabilities, $p_{\rm b}>0.9$ (independently of their starting orbital radius), 
except for the inner disk in 10/30, for which we obtain $p_{{\rm b},1} \simeq 0.85$.  The tight 
binaries are affected only during extremely close encounters to planets which do not happen 
too often. The wide binaries with $a_{\rm b}/R_{\rm b} \simeq 100$ are less likely to survive. 
Here the survival probability increases from $p_{{\rm b},1} \simeq 0.3$ for $r<30$ au
to $\simeq 0.8$--1 for bodies starting near the inner edge of the cold belt ($r=36$--42 au). 
The extreme binaries are not expected to survive (e.g., $p_{\rm b} \simeq 0.01$ for $r \lesssim 30$ au) 
unless they start with $r>37$ au. The implications of these results for blue binaries depends on 
the color transition radius $r^*$ (Fig. \ref{ratio}).

In the model, we assign color to each simulated object depending on whether it started at $r<r^*$ 
(``blue'' or LR) or $r>r^*$ (VR). The color transition at $r^*$ is assumed to be a sharp boundary 
between LR and VR. This can be a consequence of the sublimation-driven surface depletion in some 
organic molecules, such as methanol or ammonia (Brown et al. 2011). In this case, the color transition at $r^*$ 
would correspond to the sublimation radius (see Sect. 5 for a discussion).

The case with $r^* \sim 30$ au can be ruled out because the blue CC binaries would have to start 
with $r<30$ au (to have blue colors), and the great majority of wide/extreme binaries would not 
dynamically survive their implantation in the CC population. The color transition somewhere between 
30 and 42 au could be more plausible. The main constrains are: (i) to implant enough blue 
binaries in the cold disk, and (ii) not implant too many blue singles. Indeed, the existing 
observations reveal only two blue CC singles ($s<17$\%; Fraser et al. 2017, 2021). To satisfy both 
conditions, a very large binary fraction in the source region, $r<r^*$, is needed (Fraser 
et al. 2017). In addition, a large fraction
of initial binaries starting with $r<r^*$ must survive implantation; the ones that are dissolved
are a source of blue singles.

None of the options considered here seems to work to satisfy the constraints: the blue singles 
always win over the wide/extreme blue binaries. This is illustrated in Fig. \ref{ratio}. Here we assumed 
that all planetesimals are born as binaries with a fixed separation. In three different cases, we set 
$a_{\rm b}/R_{\rm b}=10$, 100, and 1000. This is not meant to represent the real situation. In 
reality, single planetesimals must have formed as well, and many initial binaries in the inner 
massive disk should have been dissociated by impacts (Nesvorn\'y \& Vokrouhlick\'y 2019; see 
Sect. 5 below), producing singles. In addition, for any reasonable assumption about their formation, 
the planetesimal binaries should have formed with a {\it range} of binary separations.

For wide binaries with $a_{\rm b}/R_{\rm b}=100$, the predicted blue-binary to blue-single number ratio, 
$R_{\rm b/s}$, is smaller than one -- in direct contradiction to observations (Fraser et al. 2017, 2021; 
$R_{\rm b/s} \simeq 4$, Table 1) -- unless $r^* > 42$ au (Fig. \ref{ratio}). This would imply that the 
blue CC binaries had to form in situ at $r=42$--47 au. We discuss this possibility in Sect. 5. 
The extreme binaries with $a_{\rm b}/R_{\rm b}=1000$ are easier to dissolve, and yield much lower values 
of $R_{\rm b/s}$. The tight binaries with $a_{\rm b}/R_{\rm b}=10$ show $R_{\rm b/s}>1$ 
for any choice of the color transition radius. This is simply because they survive and dominate the 
final statistics over singles (recall our assumption of the 100\% blue binary fraction in the source).  
    
To explain the observations that indicate the preference for blue wide/extreme binaries over tight 
ones and singles (Fraser et al. 2021), it could be imagined that all planetesimals start as tight 
binaries and these binaries become wider. This is not the case. In fact, only a very few tight 
binaries became wide/extreme our simulations -- this formation channel is simply not efficient enough. 
A similar result was obtained in Nesvorn\'y \& Vokrouhlick\'y (2019) for binaries implanted in the 
dynamically hot Kuiper belt populations. In a recent paper, Stone \& Kaib (2021) suggested that a 
small subset of tighter binaries can give rise to a small population of very wide binaries, but their 
simulations show that only $\sim1$\% of initial binaries with $30 < a_{\rm b}/R_{\rm b} < 100$ can 
evolve to have $a_{\rm b}/R_{\rm b} > 300$, with most ($\sim 90$\%) surviving with $a_{\rm b}/R_{\rm b} < 100$. 
This would not help to explain the observed statistics of blue CC binaries. 

\section{Discussion}

We end at an impasse: no simple combination of binary properties, radial distribution of planetesimals 
and color transition radius can match the constraints. Something else must be going on. We first discuss 
the influence of observational biases. An important bias that favors detection of binaries over singles 
is that a binary is composed of two objects of some size and will appear brighter than a single object
of the same size (e.g., Noll et al. 2020). For example, an exactly equal-mass binary is $\simeq 0.75$ mag 
brighter than a single object. A magnitude limited survey is therefore expected to yield a higher-than-intrinsic 
binary fraction.

A similar bias occurs when bodies are monitored for color measurements where the observational target 
must be bright enough to measure accurate colors. For the size distribution of large CCs (Fraser et al. 
2014), this could contribute by a multiplication factor of several in terms of $R_{\rm b/s}$, and is not 
large enough to resolve the issue highlighted in Fig. \ref{ratio}. In addition, the magnitude bias does 
not disfavor the detection of tight binaries (and unresolved tight binaries). We therefore find that this 
bias cannot be entirely responsible for the problem in question.\footnote{Another bias is introduced
  when CCs with unusual colors are preferentially targeted for binary-detection observations. This
  bias would favor detection of blue binaries (relative to VR binaries).}
  
In this work, we used the dynamical models from Nesvorn\'y et al. (2020). While these models were previously
shown to reproduce the orbital structure of the Kuiper belt, dynamical modeling is a source of additional
uncertainties. For example, Volk \& Malhotra (2019) showed that the mean orbital inclination of Plutionos and
HCs can be obtained in a model where Neptune migrates on a very-low-eccentricity orbit. This migration mode 
could potentially have different implications for the implantation of blue binaries (although it is not
clear, in detail, how this would work). To start with, however, it would be important to demonstrate that
the very-low-eccentricity migration of Neptune can match other constraints as well (e.g., the inclination
distribution of KBOs; see discussion in Nesvorn\'y 2021).

Levison \& Morbidelli (2003) proposed that CCs were pushed out from $\sim 35$ au to $>42$ au by the 2:1
resonance with migrating Neptune. To maintain low eccentricities in this migration model, the
resonant population was required to be massive (the total mass of several Earth masses). As bodies in the
2:1 resonance do not experience encounters with Neptune, this would favour binary survival. Moreover, if
$r^*>35$ au, the implanted binaries would be blue, potentially resolving some of the tensions discussed
above. On the down side, with several Earth masses in the 2:1 resonance, the collisional grinding in the
resonant population would presumably be intense, casting doubt on the {\it collisional} survival of wide
binaries. It is also not clear whether a very massive 2:1 population is plausible (modern
dynamical models do not place much mass beyond 30 au; e.g., Levison et al. 2008, Volk \& Malhotra 2019,
Nesvorn\'y et al. 2020).

Additional uncertainty is related to the initial inclination distribution of planetesimals. In Nesvorn\'y
et al. (2020), we adopted the Rayleigh distribution to set up the initial inclinations. For $r<30$ au,
the mean inclination was set to $\langle i \rangle = 0.06$ (3.5$^\circ$). For $r>30$ au, the mean inclination
gradually decreased with the orbital radius such that $\langle i \rangle \simeq 1.7^\circ$ for
$r=45$ au, a value comparable with the free inclination of cold classicals (Fraser et al. 2021). It is not
known what the real inclination distribution was or how it varied with the orbital radius. It is possible,
for example, that the initial inclination distribution was wider (e.g., because bodies starting in the
massive planetesimal disk have their inclinations excited by large objects that formed in the disk;
Nesvorn\'y \& Vokrouhlick\'y 2016) and more strongly varied with the orbital radius. If so, we imagine
that the implantation probability from $r \lesssim 35$ au to $42<a<47$ au, $i<5^\circ$ and $q>36$ au would
be reduced -- relative to our nominal model -- because fewer bodies could reach orbits with very low
inclinations. This would then presumably shift the balance toward having more CCs implanted from larger
orbital radii, where the binary survival is better, increase $R_{\rm b/s}$, and potentially improve the
match to observations. This is left for future work.

It is also conceivable that the wide/extreme blue binaries were not implanted in the CC population but instead 
formed in situ at $r=42$--47 au.\footnote{Fraser et al. (2021) suggested that the blue CC binaries may
  have slightly broader inclination distribution than the rest of the CC population. If confirmed, this would
  be a telltale signature of them being implanted. The statistics, with only several known blue CC binaries, did not
  allow Fraser et al. to establish this result with a high statistical significance.}
If that would be the case, however, we would need to explain why their 
LR colors differ from the predominantly VR colors of CCs. The time of formation could potentially be 
responsible for the color difference. KBOs presumably formed by the streaming instability (Youdin \& 
Goodman 2005), or a related process, in a protoplanetary gas disk. During the early stages, when the 
protoplanetary disk was massive and the solar luminosity was high (Baraffe et al. 1998), the surface 
layers of a flared disk at 45 au were heated by solar radiation and the radiation was efficiently 
diffused to the disk midplane. The midplane temperature at $r=45$ au is therefore expected to be relatively 
high (Bitsch et al. 2015). As the disk becomes less massive over time and the solar luminosity decreases, 
the temperature is expected to drop. For example, in some disk models, the midplane temperature at 45 
au changes from above 30 K during the early stages to below 20 K during the late stages (e.g., 
Bitsch et al. 2015). This would cause carbon monoxide (CO), which is the second most important gas 
molecule in astronomical environments, to freeze. 

The CO iceline near 45 au would have two important implications for the formation of CC planetesimals. 
First, the surface density of solids would be boosted near 45 au, including the contribution 
from the cold finger effect (Drazkowska \& Alibert 2017). This could potentially produce, if grains 
stick and grow to become large enough (Birnstiel et al. 2016), favorable conditions for the 
streaming instability (Carrera et al. 2015, Yang et al. 2017, Li \& Youdin 2021). 
Second, for temperatures below 30 K, CO can be destroyed on grains (chemistry mediated by cosmic ray
ionization), producing CO$_2$, CH$_3$OH (methanol), and other hydrocarbons (e.g., Bosman et al. 2018), 
all of which have relatively high sublimation temperatures and can remain on a surface. Methanol, in particular, 
has been the only ice unambiguously detected on the surface of Arrokoth (Grundy et al. 2020) -- providing 
evidence for hydrogenation of CO, possibly already on the surface of Arrokoth, but before the 
gas disk dispersal.\footnote{The average surface temperature of Arrokoth increased to $\simeq40$ K 
after the gas disk dispersal, causing rapid surface and sub-surface depletion of CO and other 
super-volatiles (Brown et al. 2011).} Methanol is known to retain high albedo and redder colors after 
irradiation (Brunetto et al. 2006), possibly the main reason behind the higher albedos and redder 
colors of CCs (Tegler \& Romanishin 2000, Brucker et al. 2009).

Blue CC binaries could have formed in situ at 45 au during the earlier protoplanetary disk stages when the 
ambient disk temperature at 45 au was higher, CO remained in the gas phase and/or the production of 
CH$_3$OH was inefficient for $T>30$ K (Bosman et al. 2018). This would lead to different chemistry and 
different composition of planetesimals formed under these conditions. The early-formed planetesimals would
subsequently be exposed to physical conditions in a late-stage protoplanetary disk (e.g., very low
temperatures), and could accrete a thin layer of methanol-rich materials (see above). This layer would
have to be later disturbed (e.g., by small impacts) to expose the interior materials. 
Our hypothesis could potentially explain the color affinity of blue CC binaries to KBOs in the dynamically
hot populations, which formed below 30 au and also experienced higher temperatures during their formation. 

The model can incorporate, after the gas disk dispersal, the implantation of bodies from $r<42$ au 
to the CC population: the majority of implanted bodies would be tight binaries and singles (Sect. 4). 
A large fraction of the implanted population -- those bodies that formed earlier and/or closer to 
the Sun -- would have LR colors. The two blue singles reported in Fraser et al. (2021) could have formed 
in-situ or been implanted. They represent $\sim 2$\% of the total CC sample with known colors 
(Fraser et al. 2021). This fraction can be used to place an upper bound on the implantation process. 
For example, we expect $\sim$10\% of the CC population to have LR colors in the 30/100 case
(for any $r^*>30$ au, in a size-limited sample; Fig. \ref{weis}). This seems excessive but note that the LR/VR ratio is 
influenced by the magnitude (see above) and albedo biases (LR objects generally have lower albedos 
than VR objects), both of which disfavor the detection of blue singles in the CC population 
(relatively to red binaries). It is therefore possible that the LR/VR ratio in the CC population is higher 
than the magnitude-limited color surveys currently indicate. 

\section{Conclusions}

Fraser et al. (2017, 2021) observationally characterized colors and binarity in the cold classical 
population of the Kuiper belt. They reported detection of eight CC binaries with LR colors (spectral slope 
$s<17$\%), most of which have very wide separations (7 of 8 have $a_{\rm b}/R_{\rm b} \gtrsim 100$), and two blue
singles (possibly unresolved binaries). The LR (or ``blue'') binaries were suggested to have been 
implanted in the CC population from $r<42$ au. Here we studied the implantation process and found:
\begin{enumerate}
\item The implantation probability increases with the increasing orbital radius of the source. 
For $r<30$ au, it is $\simeq 10^{-4}$ for fast Neptune's migration (run 10/30) and 
$\simeq 2.5 \times 10^{-5}$ for slow migration (30/100). For $r \gtrsim 38$ au, the implantation 
probability raises to $\gtrsim 10^{-3}$. 
\item Our dynamical model indicates that $\sim$10\% of cold classicals should have LR colors as these
  bodies were implanted onto CC orbits ($42<a<47$ au, $i<5^\circ$ and $q>36$ au) from below the color
  transition in the original planetesimal disk (this assumes $r^*>30$ au; Nesvorn\'y et al. 2020).
\item The blue binaries can be implanted from a wide range of initial 
heliocentric distances, but the survival of the widest binaries implies formation at $r>30$ au.
\item For any reasonable choice of parameters (Neptune's migration history, initial disk profile, 
etc.), the implantation model predicts a predominance of blue singles, rather than blue binaries, 
which contradicts existing observations. 
\end{enumerate} 
We propose that wide blue binaries formed in situ at $r=42$--47 au and their color reflects early 
formation and higher temperature of the young protoplanetary disk. The predominantly VR colors of 
cold classicals are linked to the production of methanol and other hydrocarbons during the 
late stages of the disk, when the temperature at 45 au dropped to $\simeq 20$ K and carbon monoxide 
was hydrogenated. Methanol, which is known to retain high albedo and redder colors after 
irradiation, was detected on the surface of Arrokoth.    

\acknowledgements

The work of D.N. was supported by the NASA Emerging Worlds program. The work of D.V. was 
supported by the Czech Science Foundation (grant 21-11058S).

\clearpage
\begin{table}
\centering
{
\begin{tabular}{llrrrrr}
\hline \hline
number  & designation & $R_1$ & $R_2$  &  $R_2/R_1$ &  $a_{\rm b}$  & $a_{\rm b}/R_{\rm b}$ \\  
  &                   & (km)  & (km)  &            &   (km)       &                    \\  
\hline
506121 & 2016 BP81  & 94       & 85    & 0.90 & 11,300     &  100                \\   
& 2015 RJ277        & $\sim$50 & $\sim$50 & $\sim$1       &  $\sim$1,100 & $\sim$17 \\   
511551 & 2014 UD225 & 96       & 33    & 0.34 & 21,400     &  220                \\   
& 2003 UN284        & 62       & 42    & 0.68 & 54,000     &  796                \\
& 2003 HG57         & 78       & 78    & 1.0  & 13,200     &  134                \\   
& 2002 VD131        & 107      & 27    & 0.25 & 14,900     &  139                \\   
524366 & 2001 XR254 & 86       & 70    & 0.81 &  9,310     &   94                \\
& 2001 QW322        & 64       & 63    & 0.98 & 102,100    &  1280               \\
\hline \hline
\end{tabular}
}
\caption{The blue CC binaries with spectral slope $s<17$\% as defined in Fraser et al. (2017, 2021). 
The values reported here are a mixed bag of varied-quality data. For some binaries, such as 2001 QW322 (Petit et al. 
2008), we have a detailed information about the mutual orbit. For others, such as 2015 RJ277, the 
physical radii are assumed, and the next to the last column gives the component separation observed 
at the present epoch (not the semimajor axis). Fraser et al. (2021) also reported the detection of 
two blue singles with $s<17$\% -- 1999 QE4 and 2013 GR136. The observed ratio of blue binaries to blue 
singles is therefore $R_{\rm b/s} \simeq 4$.}
\end{table}

\clearpage
\begin{table}
\centering
{
\begin{tabular}{lrrr}
\hline \hline
run id.       & $\tau_1$ & $\tau_2$   & $N_{\rm Pluto}$  \\  
                & (Myr)    & (Myr)      &            \\  
\hline
10/30            & 10       & 30          & 2000       \\
30/100           & 30       & 100         & 4000       \\
\hline \hline
\end{tabular}
}
\caption{A two stage migration of Neptune was adopted from Nesvorn\'y et al. (2020): $\tau_1$ and $\tau_2$ 
define the $e$-folding exponential migration timescales during these stages, and $N_{\rm Pluto}$ is the assumed 
number of Pluto-mass objects in the massive planetesimal disk below 30 au. 
Neptune's migration is grainy with these objects as needed 
to explain the observed proportion of resonant and non-resonant populations in the Kuiper belt (Nesvorn\'y
\& Vokrouhlick\'y 2016).}
\end{table}

\clearpage

\clearpage

\clearpage

\clearpage


\begin{thebibliography}{}

\bibitem[Agnor and Hamilton(2006)]{2006Natur.441..192A} Agnor, C.~B., Hamilton, D.~P.\ 2006.\ Neptune's 
capture of its moon Triton in a binary-planet gravitational encounter.\ Nature 441, 192-194. 

\bibitem[Ali-Dib et al.(2021)]{2021AJ....162...19A} Ali-Dib, M., Marsset, M., Wong, W.-C., et al.\ 2021, \aj, 162, 19. doi:10.3847/1538-3881/abf6ca

\bibitem[Bannister et al.(2018)]{2018ApJS..236...18B} Bannister, M.~T., Gladman, B.~J., Kavelaars, J.~J., et al.\ 2018, \apjs, 236, 1
8 

\bibitem[Baraffe et al.(1998)]{1998A&A...337..403B} Baraffe, I., Chabrier, G., Allard, F., et al.\ 1998, \aap, 337, 403

\bibitem[Birnstiel et al.(2016)]{2016SSRv..205...41B} Birnstiel, T., Fang, M., \& Johansen, A.\ 2016, \ssr, 205, 41. doi:10.1007/s11214-016-0256-1

\bibitem[Bitsch et al.(2015)]{2015A&A...575A..28B} Bitsch, B., Johansen, A., Lambrechts, M., et al.\ 2015, \aap, 575, A28. doi:10.1051/0004-6361/201424964

\bibitem[Bosman et al.(2018)]{2018A&A...618A.182B} Bosman, A.~D., Walsh, C., \& van Dishoeck, E.~F.\ 2018, \aap, 618, A182. doi:10.1051/0004-6361/201833497

\bibitem[Brown et al.(2011)]{2011ApJ...739L..60B} Brown, M.~E., Schaller, E.~L., \& Fraser, W.~C.\ 2011, \apjl, 739, L60

\bibitem[Brucker et al.(2009)]{2009Icar..201..284B} Brucker, M.~J., Grundy, W.~M., Stansberry, J.~A., et al.\ 2009, Icarus, 201, 284. doi:10.1016/j.icarus.2008.12.040

\bibitem[Brunetto et al.(2006)]{2006ApJ...644..646B} Brunetto, R., Barucci, M.~A., Dotto, E., et al.\ 2006, \apj, 644, 646. doi:10.1086/503359

\bibitem[Carrera et al.(2015)]{2015A&A...579A..43C} Carrera, D., Johansen, A., \& Davies, M.~B.\ 2015, \aap, 579, A43. doi:10.1051/0004-6361/201425120

\bibitem[Dawson \& Murray-Clay(2012)]{2012ApJ...750...43D} Dawson, R.~I. \& Murray-Clay, R.\ 2012, \apj, 750, 43. doi:10.1088/0004-637X/750/1/43

\bibitem[Dr{\k{a}}{\.z}kowska \& Alibert(2017)]{2017A&A...608A..92D} Drazkowska, J. \& Alibert, Y.\ 2017, \aap, 608, A92. doi:10.1051/0004-6361/201731491

\bibitem[Fernandez \& Ip(1981)]{1981Icar...47..470F} Fern\'andez, J.~A. \& Ip, W.-H.\ 1981, Icarus, 47, 470. doi:10.1016/0019-1035(81)90195-0

\bibitem[Fraser et al.(2014)]{2014ApJ...782..100F} Fraser, W.~C., Brown, M.~E., Morbidelli, A., Parker, A., \& Batygin, K.\ 2014, \apj, 782, 100 

\bibitem[Fraser et al.(2017)]{2017NatAs...1E..88F} Fraser, W.~C., Bannister, M.~T., Pike, R.~E., et al.\ 2017, Nature Astronomy, 1, 0088

\bibitem[Fraser et al.(2021)]{2021PSJ.....2...90F} Fraser, W.~C., Benecchi, S.~D., Kavelaars, J.~J., et al.\ 2021, PSJ, 2, 90. doi:10.3847/PSJ/abf04a

\bibitem[Grundy et al.(2020)]{2020Sci...367.3705G} Grundy, W.~M., Bird, M.~K., Britt, D.~T., et al.\ 2020, Science, 367, aay3705. doi:10.1126/science.aay3705

\bibitem[Hahn \& Malhotra(2005)]{2005AJ....130.2392H} Hahn, J.~M., \& Malhotra, R.\ 2005, \aj, 130, 2392 

\bibitem[Kavelaars et al.(2021)]{2021arXiv210706120K} Kavelaars, J., Petit, J.-M., Gladman, B., et al.\ 2021, arXiv:2107.06120

\bibitem[Levison, \& Duncan(1994)]{1994Icar..108...18L} Levison, H.~F., \& Duncan, M.~J.\ 1994, Icarus, 108, 18

\bibitem[Levison et al.(2008)]{2008Icar..196..258L} Levison, H.~F., Morbidelli, A., Van Laerhoven, C., et al.\ 2008, Icarus, 196, 258. doi:10.1016/j.icarus.2007.11.035

\bibitem[Li \& Youdin(2021)]{2021ApJ...919..107L} Li, R. \& Youdin, A.~N.\ 2021, \apj, 919, 107. doi:10.3847/1538-4357/ac0e9f

\bibitem[Malhotra(1993)]{1993Natur.365..819M} Malhotra, R.\ 1993, \nat, 365, 819. doi:10.1038/365819a0

\bibitem[Morbidelli \& Nesvorn{\'y}(2020)]{2020tnss.book...25M} Morbidelli, A. \& Nesvorn{\'y}, D.\ 2020, The Trans-Neptunian Solar System, 25. doi:10.1016/B978-0-12-816490-7.00002-3

\bibitem[Nesvorn{\'y}(2015)]{2015AJ....150...73N} Nesvorn{\'y}, D.\ 2015a, \aj, 150, 73 

\bibitem[Nesvorn{\'y}(2015)]{2015AJ....150...68N} Nesvorn{\'y}, D.\ 2015b, \aj, 150, 68 

\bibitem[Nesvorn{\'y}(2021)]{2021ApJ...908L..47N} Nesvorn{\'y}, D.\ 2021, \apjl, 908, L47. doi:10.3847/2041-8213/abe38f

\bibitem[Nesvorn{\'y} \& Morbidelli(2012)]{2012AJ....144..117N} Nesvorn{\'y}, D., \& Morbidelli, A.\ 2012 (NM12), \aj, 144, 117 

\bibitem[Nesvorn{\'y} \& Vokrouhlick{\'y}(2016)]{2016ApJ...825...94N} Nesvorn{\'y}, D., \& Vokrouhlick{\'y}, D.\ 2016, \apj, 825, 94 

\bibitem[Nesvorn{\'y}, \& Vokrouhlick{\'y}(2019)]{2019Icar..331...49N} Nesvorn{\'y}, D., \& Vokrouhlick{\'y}, D.\ 2019, Icarus, 331, 49

\bibitem[Nesvorn{\'y} et al.(2010)]{2010AJ....140..785N} Nesvorn{\'y}, D., Youdin, A.~N., \& Richardson, D.~C.\ 2010, \aj, 140, 785. doi:10.1088/0004-6256/140/3/785

\bibitem[Nesvorn{\'y} et al.(2019)]{2019NatAs...3..808N} Nesvorn{\'y}, D., Li, R., Youdin, A.~N., et al.\ 2019, Nature Astronomy, 3, 808. doi:10.1038/s41550-019-0806-z

\bibitem[Nesvorn{\'y} et al.(2020)]{2020AJ....160...46N} Nesvorn{\'y}, D., Vokrouhlick{\'y}, D., Alexandersen, M., et al.\ 2020, \aj, 160, 46. doi:10.3847/1538-3881/ab98fb

\bibitem[Noll et al.(2020)]{2020tnss.book..201N} Noll, K., Grundy, W.~M., Nesvorn{\'y}, D., et al.\ 2020, The Trans-Neptunian Solar System, 201. doi:10.1016/B978-0-12-816490-7.00009-6

\bibitem[Parker \& Kavelaars(2010)]{2010ApJ...722L.204P} Parker, A.~H. \& Kavelaars, J.~J.\ 2010, \apjl, 722, L204. doi:10.1088/2041-8205/722/2/L204

\bibitem[Petit \& Mousis(2004)]{2004Icar..168..409P} Petit, J.-M. \& Mousis, O.\ 2004, Icarus, 168, 409. doi:10.1016/j.icarus.2003.12.013

\bibitem[Petit et al.(2008)]{2008Sci...322..432P} Petit, J.-M., Kavelaars, J.~J., Gladman, B.~J., et al.\ 2008, Science, 322, 432. doi:10.1126/science.1163148

\bibitem[Press et al.(1992)]{1992nrfa.book.....P} Press, W.~H., Teukolsky, S.~A., Vetterling, W.~T., et al.\ 1992, Cambridge: University Press, 2nd edition

\bibitem[Stone \& Kaib(2021)]{2021MNRAS.505L..31S} Stone, L.~R. \& Kaib, N.~A.\ 2021, \mnras, 505, L31. doi:10.1093/mnrasl/slab044

\bibitem[Tegler and Romanishin(1998)]{1998Natur.392...49T} Tegler, S.~C., Romanishin, W.\ 1998, Nature 392, 49 

\bibitem[Tsiganis et al.(2005)]{2005Natur.435..459T} Tsiganis, K., Gomes, R., Morbidelli, A., et al.\ 2005, \nat, 435, 459. doi:10.1038/nature03539

\bibitem[Volk \& Malhotra(2019)]{2019AJ....158...64V} Volk, K. \& Malhotra, R.\ 2019, \aj, 158, 64. doi:10.3847/1538-3881/ab2639
  
\bibitem[Wong, \& Brown(2017)]{2017AJ....153..145W} Wong, I., \& Brown, M.~E.\ 2017, \aj, 153, 145

\bibitem[Yang et al.(2017)]{2017A&A...606A..80Y} Yang, C.-C., Johansen, A., \& Carrera, D.\ 2017, \aap, 606, A80. doi:10.1051/0004-6361/201630106

\bibitem[Youdin, \& Goodman(2005)]{2005ApJ...620..459Y} Youdin, A.~N., \& Goodman, J.\ 2005, \apj, 620, 459


\end{thebibliography}
\end{document}